\documentclass[conference]{IEEEtran}
\IEEEoverridecommandlockouts
\usepackage{cite}
\usepackage{amsmath,amssymb,amsfonts}
\usepackage{mathtools}

\usepackage{amsmath}
\usepackage{algorithmic}
\usepackage{graphicx}
\usepackage{quantikz}
\usepackage{caption}
\usepackage{subcaption}
\usepackage{textcomp}
\usepackage{xcolor}
\usepackage{float}
\usepackage{listings}
\usepackage{xcolor}
\usepackage{braket}
\usepackage[colorlinks=true,allcolors={blue}]{hyperref}

\definecolor{codegreen}{rgb}{0,0.6,0}
\definecolor{codegray}{rgb}{0.5,0.5,0.5}
\definecolor{codepurple}{rgb}{0.58,0,0.82}
\definecolor{backcolour}{rgb}{0.95,0.95,0.92}

\lstdefinestyle{mystyle}{
    backgroundcolor=\color{backcolour},   
    commentstyle=\color{codegreen},
    keywordstyle=\color{magenta},
    numberstyle=\tiny\color{codegray},
    stringstyle=\color{codepurple},
    basicstyle=\ttfamily\footnotesize,
    breakatwhitespace=false,         
    breaklines=true,                 
    captionpos=b,                    
    keepspaces=true,                 
    numbers=left,                    
    numbersep=5pt,                  
    showspaces=false,                
    showstringspaces=false,
    showtabs=false,                  
    tabsize=2
}

\def\BibTeX{{\rm B\kern-.05em{\sc i\kern-.025em b}\kern-.08em
    T\kern-.1667em\lower.7ex\hbox{E}\kern-.125emX}}
\begin{document}

\title{Quantum Machine Learning: Fad or Future?\\
}

\author{\IEEEauthorblockN{1\textsuperscript{st} Arhum Ishtiaq}
\IEEEauthorblockA{\textit{Computer Science Department} \\
\textit{Habib University}\\
Karachi, Pakistan \\
ai05182@st.habib,edu.pk}
\and

\IEEEauthorblockN{2\textsuperscript{nd} Sara Mahmood}
\IEEEauthorblockA{\textit{Computer Science Department} \\
\textit{Habib University}\\
Karachi, Pakistan \\
sm05155@st.habib,edu.pk}

}

\maketitle

\begin{abstract}
For the last few decades, classical machine learning has allowed us to improve the lives of many through automation, natural language processing, predictive analytics and much more. However, a major concern is the fact that we're fast approach the threshold of the maximum possible computational capacity available to us by the means of classical computing devices including CPUs, GPUs and Application Specific Integrated Circuits (ASICs). This is due to the exponential increase in model sizes which now have parameters in the magnitude of billions and trillions, requiring a significant amount of computing resources across a significant amount of time, just to converge one single model. To observe the efficacy of using quantum computing for certain machine learning tasks and explore the improved potential of convergence, error reduction and robustness to noisy data, this paper will look forth to test and verify the aspects in which quantum machine learning can help improve over classical machine learning approaches while also shedding light on the likely limitations that have prevented quantum approaches to become the mainstream. A major focus will be to recreate the work by Farhi et al. in \cite{b1} and conduct experiments using their theory of performing machine learning in a quantum context, with assistance from the \href{https://www.tensorflow.org/quantum/}{Tensorflow Quantum documentation}.
\end{abstract}


\section{Introduction}

The primary experiment conducted for this paper will focus on creating three neural networks with the goal of being able to recognize handwritten digits from the MNIST dataset, which comprises of 70,000 images of handwritten digits with a resolution of 28x28. The first neural network will be an 8-layer CNN with nearly 1.2 million parameters.This model will represent the state-of-the-art in current classical approaches of machine learning. Secondly, 3 versions of a single-layer quantum neural network will be created with a selection of 8, 18, and 32 parameters. The single layer is actually created through Tensorflow's Parametrized Quantum Circuit (PQC) function which creates specialized layer that allows us to train our model circuit using the quantum data we provide. Finally, the third neural network will be roughly equivalent to our quantum neural networks in terms of the total parameters in the model, which in this case will be 13, 23, 37, across 3 layers. The visualizations of the model architecture we'll use across this experiment are as follows:

\begin{figure}[H]
\centering
\begin{subfigure}[t]{0.3\linewidth}
\centering
\includegraphics[width=\textwidth]{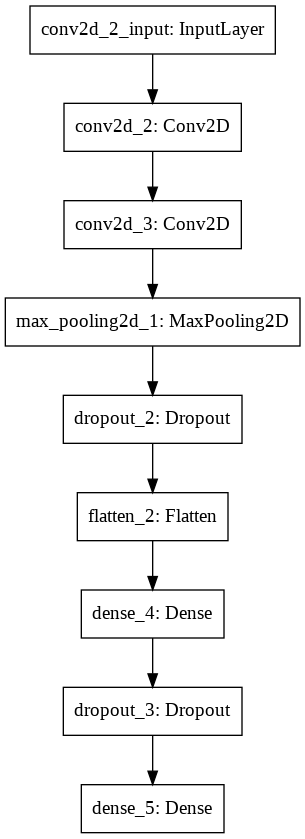}
\caption{Full-sized network architecture}\label{fig1a}
\end{subfigure}%
~
\begin{subfigure}[t]{0.3\linewidth}
\centering
\includegraphics[width=\textwidth]{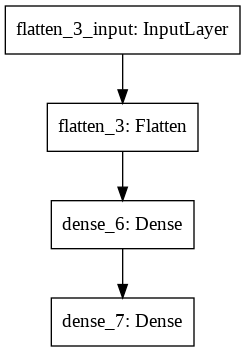}
\caption{Fair-sized network architecture}\label{fig1b}
\end{subfigure}%
~
\begin{subfigure}[t]{0.25\linewidth}
\centering
\includegraphics[width=\textwidth]{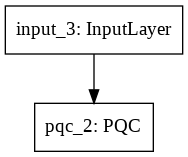}
\caption{Quantum network architecture}\label{fig1c}
\end{subfigure}%

\caption{Model Architecture Visualization}

\end{figure}

Before moving on to the details of the experiment and the results, it is imperative that we establish an understanding on how classical machine learning can be translated into quantum machine learning as it requires a completely different problem representation than what is standard in the current machine learning domain. Therefore, we will go on to provide details regarding the gates and operators that will be used, along with an overview of the translation of classical to quantum machine learning given a supervised learning problem. We will also discuss the model architectures and data representations that will be implemented as part of our experiment.

\section{Background}

In this section we'll discuss the overview and definitions of some of the quantum gates which will aid in understanding the overall report.
\subsection{Pauli Operators}\label{2.1}
Pauli operators are logical operators or logical gates which when applied to input state allows us to express the effects of the environment on a qubit of the states \cite{b2}. In  this implementation Pauli gates are  used to make up the neural network model circuit. There are in total 4 different kinds of Pauli operators.
\subsubsection{Definitions}
\begin{enumerate}
    
    \item \textbf{X gate} \\
    X gate is a bit flip operation and changes the bit from 0 to 1 and vice versa. Representation of the X gate is as follows:\\
    \begin{center}
   $\begin{pmatrix}
       0 & 1\\
       1& 0 
    \end{pmatrix}$
    \begin{equation}
        \ket{\psi} = \alpha_{1}\ket{0}  +\alpha_{0}\ket{1}
    \end{equation}

    \end{center}
   \begin{table}[H]
    \centering
    \begin{tabular}{|c|c|}
    \hline
    \textbf{Input} & \textbf{Output} \\
    \hline
    0 & 1\\
    \hline
     1 & 0\\
    \hline
    \end{tabular}
    \caption{Truth table for X gate}
    \label{tab:my_label}
\end{table}
\begin{center}
\begin{figure}
    \centering
    \includegraphics[width=50mm]{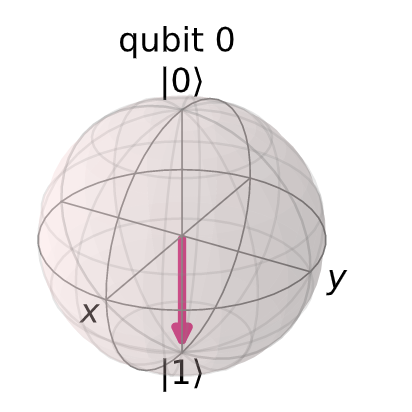}
    \caption{Bloch sphere}
    \label{fig:my_label}
\end{figure}

\end{center}
If we consider the effect X gate on a bloch sphere instead of a matrix we see that it is actually a rotation of $\pi$ around the sphere \cite{b2}.

    \item \textbf{Z gate} \\
    Z gate is a phase flip operation. It rotates the input by $\pi$ around the z axis on the bloch sphere \cite{b2}. Z gate is  a special type of phase shift gate discussed in sec. \ref{2.2}. Representation of the z gate is as follows:\\
    
\begin{center}

      $ \begin{pmatrix}
       1 & 0\\
       0& -1
    \end{pmatrix}$
\begin{equation}
          Z\ket{\psi} = \alpha_{0}\ket{0}   - \alpha_{1}\ket{1}  
\end{equation}

    \begin{table}[H]
    \centering
    \begin{tabular}{|c|c|}
    \hline
        \textbf{Input} & \textbf{Output} \\
    \hline
        $\ket{0}$ & $\ket{0}$\\
    \hline
         $\ket{1}$ & -$\ket{1}$\\
    \hline
    \end{tabular}
    \caption{Truth table for Z gate}
    \label{tab:my_label}
\end{table}
\end{center}

\end{enumerate}

\subsection{Phase Shift Gate}\label{2.2}
Phase shift gate are the gates that map $ \ket{0}$ to $\ket{0}$ and $\ket{1}$ to $ e^{\iota \varphi}\ket{1} $. The probability of the outcome does not change because of this but the phase of the quantum state is shifted. On the bloch sphere it is equivalent to tracing a latitudinal line of radian $\varphi$ \cite{b3}. The matrix representation for these functions are as follows:
\begin{equation}
    P(\varphi) = 
    \begin{pmatrix}
       1& 0\\
       0& e^{\iota \varphi}
    \end{pmatrix}
\end{equation}

\subsection{Hadamard Gate}
It is a gate which maps $\ket{0}$ to $\frac{\ket{0}+ \ket{1}}{\sqrt{2}}$ and maps $\ket{1}$ to $\frac{\ket{0}- \ket{1}}{\sqrt{2}}$ this tells us that after applying Hadamard the measurement we get has an equal probability of being equal to 0 or 1 which means this gate is used for creating superpositions. This gate is a combination of two rotation. First one being the rotation of $\pi$ around the z axis and then $\frac{\pi}{2}$ around the y axis \cite{b4}.
The matrix representation for the gate goes as follows:
\begin{equation}
    H = \frac{1}{\sqrt{2}}\begin{pmatrix}
        1& 1\\
       1& -1
    \end{pmatrix}
\end{equation}

\section{Classical to Quantum Conversion}
In this section and a few sections after this, we will discuss in detail how the classical machine learning problem of classification through supervised learning is translated into a quantum machine learning algorithm. The notations and concepts described in these sections are in accordance to the algorithm proposed in \cite{b5}. To understand how the given problem is translated into a quantum circuit let us first understand the data representation. Data is represented as set of strings where each string consists of $ z= z_1,z_2...$ where each $z_i$ is a bit of value +1 or -1. with each data point there is a label attached $l(z) = +1/-1$.In theory it is safe to assume that the data sets consists of all $2^n$ number of strings. To quantum processors act on n+1 qubits and the last bit acts as the readout bit. A unitary transformation is applied on each of the input state which depends on the parameter $\theta$. This is the parameter that our circuit needs to learn for each input state. The unitary transformation is represented as $U(\theta)$. For this implementation we say that each unitary function is of the form:
\begin{equation}
    U(\theta) = exp(\iota \theta_{i} \Sigma)
\end{equation}

Here the $\Sigma$ represents a generalised Pauli operator among the last three Pauli gates described in sec. \ref{2.1}. It is not necessary that $\Sigma$ represents a single operator but in some scenarios can represent a combination of these gates. In this case it can also be said that it is a tensor product of the Pauli gates acting on the qubits.

The above unitary function acts on a subset of qubits and for the sake of simplicity the unitary function only depends on one parameter $ \theta$. 

We take a combination of unitary functions of length L such that:
\begin{equation}
    U(\overrightarrow{\theta)} = U_{L}(\theta_{L})U_{L-1}(\theta_{L-1})....U_{1}(\theta_{1})
\end{equation}

The above function depends on $\overrightarrow{L}$  parameters. Moving on for each a computational basis states are constructed such that
\begin{equation}
    \ket{z,1}  =  \ket{z_{1}z_{2}.....z_{n}, 1}
\end{equation}

The readout bit in the above state is set to 1.
After applying the unitary function on the above states the output is as follows:
\begin{equation}\label{u}
    U(\overrightarrow{\theta})  = \ket{z,1}
\end{equation}

 We then apply a Pauli operator $\sigma_{y}$, which we decide according to our implementation and are explained in detail in sec. \ref{2.1}, on the readout qubit. We denote the operator as $Y_{n+1}$.  This operation gives us +1 or -1 as an output label and we aim to make them in accordance with the actual label which is $l(z)$. The result of $Y_{n+1}$ is not definite rather a real number between -1 and +1 and the equation for the procedure is represented as follows:
\begin{equation}
    \Bra{z,1}|U^\dag(\overrightarrow{\theta}) Y_{n+1} U(\overrightarrow{\theta})\ket{z,1}
\end{equation}
The output is a real number since it is the average of the outputs of $Y_{n+1}$ when it is measured accross multiple copies of Eq. \ref{u}
Our goal is to learn $\overrightarrow{\theta}$ such that the outcome of the above function is near the actual label. For that we also calculate the loss through the following loss function. 
\begin{equation}
    loss(\overrightarrow{\theta} , 1) = 1 - l(z)       \Bra{z,1}|U^\dag(\overrightarrow{\theta}) Y_{n+1} U(\overrightarrow{\theta})\ket{z,1} \label{eq:12}
\end{equation}

This will return 0 if the network is working perfectly and has no error and is predicting all the labels correctly. If the network is predicting randomly then the maximum loss is 1.

\begin{figure}[H]

\centering
\includegraphics[width=0.5\textwidth]{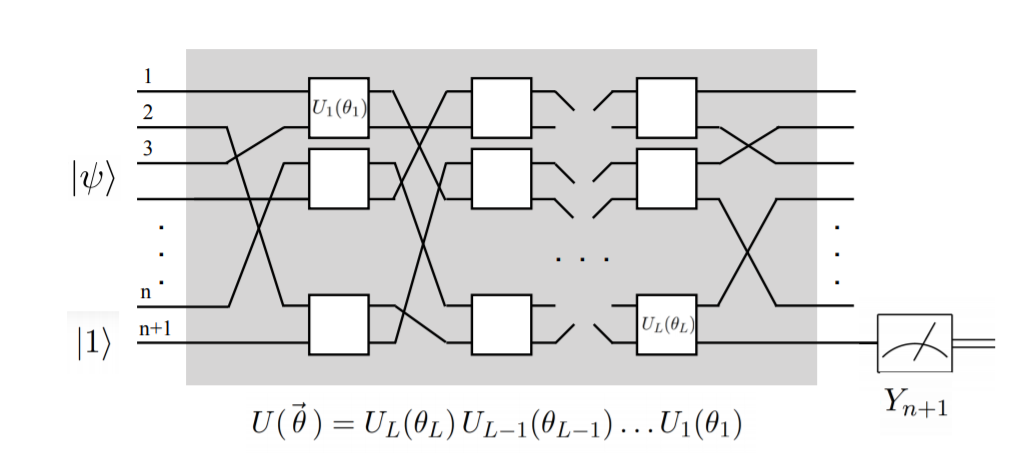}

\caption{Quantum Neural Network On a Quantum Processor}
\label{fig:figure3}

\end{figure}

Figure \ref{fig:figure3} represents the proposed quantum neural network on a quantum processors. The figure shows that a quantum state $\ket{\psi, 1}$ is passed as input and transformed through a series of unitary functions $U_{i}(\theta_{i})$. These $\theta$ are the parameters that are learnt during the learning process explained in sec. \ref{lr} such that the readout bit $Y_{n+1}$ is the desired label for the input.

\section{Supervised Learning of Parameters in Quantum Computing}\label{lr}
The process of learning the optimal parameters is the same as that in the classical machine learning process that is ‘Stochastic Gradient Descent’\cite{b6}. In this process we initialize a parameter $\overrightarrow{\theta}$ randomly and take a single sample set $z^1$ from the training dataset. We calculate the prediction by $U(\overrightarrow{\theta})\ket{z^1,1}$ and measure the chosen Pauli operator on the readout bit. Then calculate the loss from the loss function described in Eq. \eqref{eq:12}, We make small changes in the $\overrightarrow{\theta}$ to reduce its loss on the chosen sample $z^1$. Although one way to do this is to randomly sample from nearby $\overrightarrow{\theta}$ but the optimal way is to calculate the gradient w.r.t $\overrightarrow{\theta}$ of the loss and take a small step in this direction to minimize the loss. The new $\overrightarrow{\theta}$ is then used to make predictions on the next sample $z^2$ and the loss of the prediction made on $z^2$ is calculated to further update $\overrightarrow{\theta}$ the whole process is repeated till $\overrightarrow{\theta}$ is learnt that can make accurate prediction when applied to any given sample.

It is important to understand the way gradient is calculated on the loss function during the process of stochastic gradient descent method.  To take the gradient we differentiate Eq. \eqref{eq:12} with respect to $\theta$. Since the unitary function $U(\overrightarrow{\theta}$) contains L number of uniary functions each dependant on a single $\theta_i$. We calculate the gradient on Eq. \eqref{eq:12} with respect to each $\theta$ one by one. For a unitary function $U_{k}(\theta_{k}$ containing a generalized Pauli operator $\Sigma_{k}$ the gradient of the loss function becomes:\\
Let,
\begin{equation}\label{eq:13}
    \Bra{z,1}U^{\dag}_{1}..U^{\dag}_{L}Y_{n+1}U_{L}..U_{k+1}\Sigma_{k}U_{k}...U_{1}\ket{z,1}
\end{equation}

\begin{equation}\label{diff}
    \frac{dloss(\overrightarrow{\theta} , z)}{d\theta_{k}} = 2imag(Eq. \eqref{eq:13})
\end{equation}
This means the differentiation of the loss function Eq. \eqref{eq:12} is twice the imaginary part of Eq. \eqref{eq:13}, hence the term was enclosed in imag() bracket. This notation will be used to refer to the imaginary part of the equation throughout the report. Here the $Y_{n+1}$ and the $\Sigma_{k}$ both are Pauli operators forming the unitary operators. We can define the combined unitary operator as:
\begin{equation}
    \mathcal{U} = U^{\dag}_{1}..U^{\dag}_{L}Y_{n+1}U_{L}..U_{k+1}\Sigma_{k}U_{k}...U_{1}
\end{equation}
So equation 3 can now be written as:
\begin{equation}
    \frac{loss(\overrightarrow{\theta} , z)}{d\theta_{z}} =2imag(\Bra{z,1}\mathcal{U}\ket{z,1})
\end{equation}
To measure the right hand side of the above equation the $\mathcal{U}$ acts on $\ket{z,1}$ using an auxiliary qubit. The process starts as follows:
\begin{equation}
    \ket{z,1} \frac{1}{2}(\ket{0} + \ket{1})
\end{equation}
The auxiliary qubit is conditioned to be 1 and therefore when $\iota\mathcal{U}(\overrightarrow{\theta})$ acts on the above equation it becomes:
\begin{equation}
      \frac{1}{\sqrt{2}}(\ket{z,1} \ket{0} + \iota\mathcal{U}(\overrightarrow{\theta})\ket{z,1} \ket{1})
\end{equation}
Applying Hadamard on the auxilary qubit we get:
\begin{equation}
      \frac{1}{\sqrt{2}}(\ket{z,1} + \iota\mathcal{U}(\overrightarrow{\theta})\ket{z,1} \ket{1}) +       \frac{1}{2}(\ket{z,1} - \iota\mathcal{U}(\overrightarrow{\theta})\ket{z,1} \ket{1})
\end{equation}
After measuring auxiliary qubit from the above equation the probability to get 0 is 
\begin{equation}
    \frac{1}{2} - \frac{1}{2} imag(\Bra{z,1}\mathcal{U}\ket{z,1})
\end{equation}
We repeat this process multiple times to make a good estimate of the imaginary part which in turn becomes the estimate of $k_{th}$ component of the gradient. 
Once an optimal estimate gradient is found we update the $\overrightarrow{\theta}$ according to the following strategy:
\begin{equation}
    \overrightarrow{\theta} \rightarrow  \overrightarrow{\theta}- r\left(\frac{loss(\overrightarrow{\theta} , z)}{\overrightarrow{g}^2}\right) \overrightarrow{g}
\end{equation}
The strategy applied above is similar to the one used in classical machine learning. Here r is the learning rate which is a small number and is decided and adjusted to get optimal results and $\overrightarrow{g}$ is the direction of the gradient in which the $\overrightarrow{\theta}$ is updated. The introduction of this learning rate ensures that the $\overrightarrow{\theta}$ is updated in a manner that it although improves the results on the current sample but does not make the predictions worse for other samples either.
 
\section{Alternate Data Representation}
The main goal for the learning process as discussed in sec. \ref{lr} is that  parameter $\overrightarrow{\theta}$ is learnt for the unitary function such that given a n-qubit state and the readout bit set to 1 the function returns $Y_{n+1}$ such that it is the appropriate label. For this purpose so far we've talked about representing each data point separately but in \cite{b5} an alternate representation of classical data is proposed which is in the form of superpositions. We do not utilize this approach in our implementation explained in the sec. \ref{exp}  but it is worth while to take a look and see how we can represent all the data points having the same label as a single superposition. For this representation data is divided into 2 superpositions one is represented as $\ket{+1}$ and the other one is represented as $\ket{-1}$  since the focus is on binary classification this suffices. Each state can be seen as a batch that contains all the data of the same labels. 
\begin{equation}
    \ket{+1} = N_{+} \sum_{z:l(z) = 1} e^{\iota \varphi_{z}} \ket{z,1}
\end{equation}
\begin{equation}
    \ket{-1} = N_{-} \sum_{z:l(z) = -1} e^{\iota \varphi_{z}} \ket{z,1}
\end{equation}
Here the N terms are the normalization factors and since the phase does not really matter as a phase change would not affect the probability of the outcome as we discusses in sec. \ref{2.2} in detail, we can keep the phase $\varphi$ as 0 for both.

 The major difference in this approach from the one in sec. \ref{lr} is that we give a superposition $\ket{\psi}$ as input rather than a  data string z. And the readout bit is again set to 1. The unitary state becomes:
\begin{equation}
    U(\overrightarrow{\theta})\ket{\psi ,1}
\end{equation}
due to this difference a major change takes place in the loss function, Eq. \eqref{eq:12}. When the unitary function $\overrightarrow{\theta}$ is applied to the input state $\ket{-1}$ the expected value of $Y_{n+1}$ is the average of all the samples of the that  have the label -1 of the quantum neural network and same is the case with samples having +1 labels. Therefore the loss function which is optimized becomes:
\begin{equation}
    1 - \frac{1}{2}(\Bra{+1}U^{\dag}(\overrightarrow{\theta})Y_{n+1}U(\overrightarrow{\theta}\ket{+1} - \Bra{-1}U^{\dag}(\overrightarrow{\theta})Y_{n+1}U(\overrightarrow{\theta})\ket{-1})
\end{equation}

\section{Experiment}\label{exp}
\subsection{Data Preprocessing}
To implement the aforementioned theory in section 3 and 4, we first had to process the MNIST training data and make it binary. This meant that the most optimal and fastest approach was to pick any two of the ten class labels, represent them as either True and False values, and use that data as the input to the QNN, as shown in the function below.

\begin{lstlisting}[language=Python]
def filter(label_1, label_2, x, y):
    keep = (y == label_1) | (y == label_2)
    x, y = x[keep], y[keep]
    y = y == label_1
    return x,y
\end{lstlisting}

For the sake of our experiment we chose the labels of 3 and 6, however any other arbitrary labels could be chosen instead of them. After this, we had to address the first big limitation of quantum machine learning: the lack of general availability of quantum computational capacity. This meant that the image size of the MNIST dataset had to be downsampled from 28x28 to three potential values that we tested for i.e. 4x4, 3x3 and 2x2. After downsampling, we needed to convert the image itself to a circuit, which was done with the following code:
 
\begin{lstlisting}[language=Python]
def convert_to_circuit(image):
    # Encode downscaled image to a quantum datapoint
    values = np.ndarray.flatten(image)
    qubits = cirq.GridQubit.rect(DIMENSIONS, DIMENSIONS)
    circuit = cirq.Circuit()
    for i, value in enumerate(values):
        if value:
            circuit.append(cirq.X(qubits[i]))
    return circuit
\end{lstlisting}

\subsection{Model Architecture}
For the architecture of our QNN model, we decided to apply the unitary functions/gates on two qubits at a time, as suggested in \cite{b5}. One of the two qubits is the readout bit while the other one is any of the data qubits. From section 8 we can see that for our implementation the number of data qubits varies from 4x4 to 3x3 and 2x2. First we apply the X gate and the Hadamard gate on the readout qubit and then add a layer of XX and ZZ layer which is equivalent to applying X and Z gate on both the input qubits \cite{b7}. Finally we apply a layer of Hadamard  on the readout qubit. For 17-qubit input where the data qubits are 4x4 and the readout qubit is set to be 1, the number of parameters are equal to $2* 16 = 32$. Similarly, for a 10-qubit input there are $2*9 = 18$ parameters and for 5-qubit input we get $2*4 = 8$ parameters. Following is the implementation of this architecture in code:

\begin{lstlisting}[language=Python]
def create_quantum_model(dim):
    # dim = [2,3,4]
    data_qubits = cirq.GridQubit.rect(dim, dim) 
    # readout is a single qubit at [-1,-1]
    readout = cirq.GridQubit(-1, -1) 
    circuit = cirq.Circuit()
    
    # Prepare the readout qubit
    circuit.append(cirq.X(readout))
    circuit.append(cirq.H(readout))
    bldr = CircuitLayerBuilder(
        data_qubits = data_qubits,
        readout=readout)

    # Add layers
    bldr.add_layer(circuit,cirq.XX,"xx1")
    bldr.add_layer(circuit,cirq.ZZ,"zz1")

    # Finally, prepare the readout qubit.
    circuit.append(cirq.H(readout))

    return circuit, cirq.Z(readout)
\end{lstlisting}

\section{Results}
\subsection{Best QNN hyperparameters}
To better understand the efficacy of a QNN over a classical NN, we first need to establish a baseline for what a competent QNN is. This can be defined in terms of finding the QNN with the best possible accuracy metric and then finding out what particular hyperparameter combination made that performance possible. Therefore, the first analysis of our result will be looking at accuracies of our 6 distinct QNNs and find a trend that allows us to extrapolate the optimal direction we need to take while tuning our hyperparameters. 

\begin{figure}[H]
    \centering
    \includegraphics[width=0.5\textwidth]{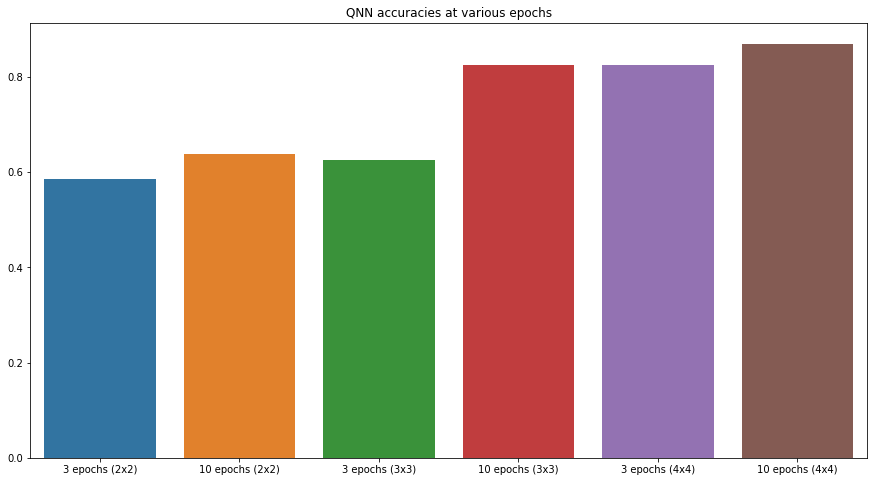}
    \caption{Comparison of QNN variations based on changes in epochs and input size}
    \label{fig:my_label3}
\end{figure}

\begin{figure}[H]
    \centering
    \includegraphics[width=0.5\textwidth]{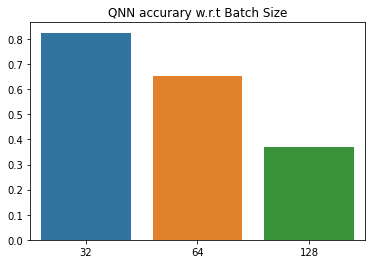}
    \caption{Comparison of QNN variations based on changes in training batch size}
    \label{fig:my_label4}
\end{figure}

As we can see, the primary driver for an increase in performance is the overall dimension of our input size. This entails that with an increased resolution, we can improve our performance significantly. However, a major hurdle to realizing that increase in performance is the exponential increase in computation required to handle a dimension increase of 1 pixel for each axis. While a QNN with input size 4x4 could process an epoch in one minute, increasing the input size to 5x5 increased the average epoch time to 90 minutes. Which brings us to epochs; we can see that while increasing the total epochs helps improve the model performance, the trade off with regards to the increased training time may not be worth it, unless in the case of an input size of 3x3 where our model performance improved by nearly 50\%. Next up, we try to understand the effect of batch size on the overall performance of the QNN, depicted in Figure \ref{fig:my_label4}. 

Here we can see that by doubling our batch size, we lose nearly one-third of our model accuracy performance. We believe that this is due to the fact that, with a smaller batch size, the model gets to learn over much more steps over a given epoch than it would with a larger batch size. Next up, we take a look at the effect of input dimensions on the models' performance metrics.

\begin{figure}[H]
    \centering
    \includegraphics[width=0.5\textwidth]{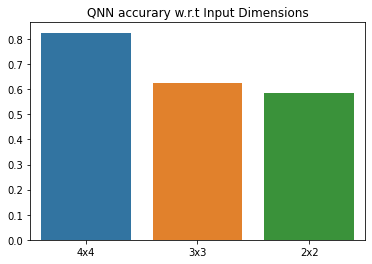}
    \caption{Comparison of QNN variations based on changes in input sizes}
    \label{fig:my_label5}
\end{figure}

Based on the graph above, we can empirically see that increased input dimensions allow for increased overall model accuracy but what is interesting is the accuracy delta between each of the input dimensions. While 4x4 allows for the best performance, the difference between the performance of 3x3 and 2x2 resolutions is very insignificant. This could be most likely due to how the downsampling process represented the data such that the models could not learn much from inputs of size 3x3 than they could from those of size 2x2.

\subsection{Comparative Analysis}
\begin{figure}[H]
    \centering
    \includegraphics[width=0.5\textwidth]{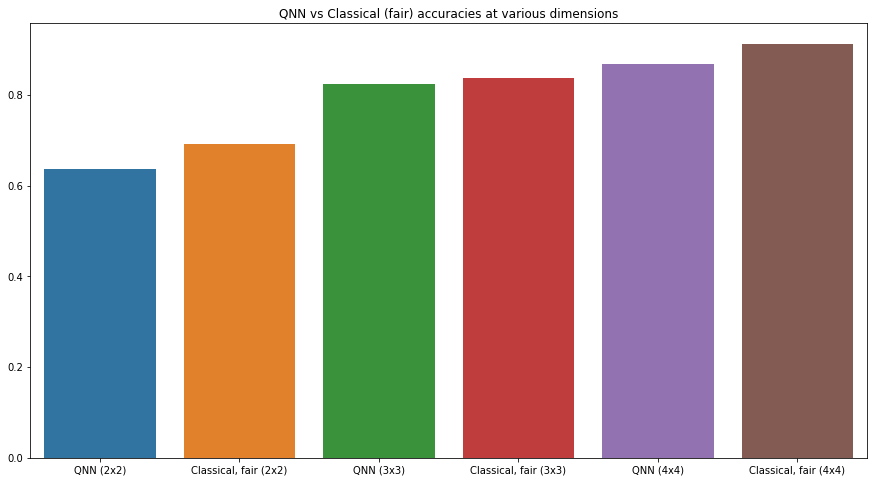}
    \caption{Performance of QNN and classical (fair) models}
    \label{fig:my_label6}
\end{figure}

Finally, we will now directly compare the results of our QNN and the classical (fair) model with regards to their performance accuracy at various input sizes. Figure \ref{fig:my_label6} shows the unfortunate reality that even after catering to reducing the input size, model complexity, number of parameters, and training epochs, the QNN does not surpass the performance capability of even the fairly-designed one, let alone the fully-sized one. While it is to be noted that performance delta that is much less than 10\%, we can empirically see that QNN does not measure up to the classical neural network approach. 

\begin{figure}[H]
    \centering
    \includegraphics[width=0.5\textwidth]{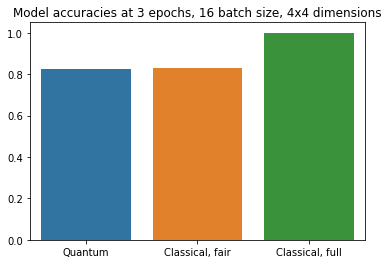}
    \caption{Model Comparison}
    \label{fig:my_label7}
\end{figure}

There was only one set of hyperparameters where the performance between the QNN and the fair model was the same and it is shown in Figure \ref{fig:my_label7}, where the QNN had the advantage of a really small batch size, highest possible input size and a low epoch that hindered the convergence of the fair model.

\section{Conclusion}
Through this experiment, we established the prominence of classical machine learning, discussed the potential obstacles that we may face in the future and whether quantum machine learning could help replace the conventional approaches along with conducting an experiment to confirm or deny the aforementioned claim. We also went on to discus the theory behind how a classical neural network can be converted to a quantum neural network by representing the layers using unitary functions and changing our input from an image to a circuit representation as well. To test and compare our conversion with the classical versions, we conducted a thorough experiment that went on to showcase the unfortunate reality that quantum machine learning still has a ways to go. The primary obstacles included the limited computational capabilities in terms of being able to process data that is magnitudes smaller in resolution/detail (compared to even some of the basic classical implementations out there) along with a significant lack of software support and infrastructure that classical machine learning really benefits from in terms of optimization, libraries etc. A part of the experiment also focused on how various hyperparameters (epochs, input sizes, models sizes, and batch sizes) affect the performance of any given quantum neural network and we concluded the fact that the only QNN that can compete with a classical (fair) model would be a result of training for 3 epochs with a batch size of 16 and input size of 4x4. Thus, it is very evident that quantum machine learning is in a very nascent stage and for it to become useful enough to replace the classical approach, there has to be a lot of work done in developing both hardware and software capabilities that allow for increased efficiency, data processing, and parallelization. While the results may not have proven to be hopeful to any extent, there is certainly a lot of potential in quantum machine learning serving as an alternative methodology of making machine smarter.

\end{document}